# Experimental Extraction and Simulation of Charge Trapping during Endurance of FeFET with TiN/HfZrO/SiO$_2$/Si (MFIS) Gate Structure

Shujing Zhao, Fengbin Tian, Hao Xu, Jinjuan Xiang, Tingting Li, Junshuai Chai, Jiahui Duan, Kai Han, Xiaolei Wang, Wenwu Wang, and Tianchun Ye

*Abstract*—We investigate the charge trapping during endurance fatigue of FeFET with TiN/Hf$_{0.5}$Zr$_{0.5}$O$_2$/SiO$_2$/Si (MFIS) gate structure. We propose a method of experimentally extracting the number of trapped charges during the memory operation, by measuring the charges in the metal gate and Si substrate. We verify that the amount of trapped charges increases during the endurance fatigue process. This is the first time that the trapped charges are directly experimentally extracted and verified to increase during endurance fatigue. Moreover, we model the interplay between the trapped charges and ferroelectric polarization switching during endurance fatigue. Through the consistency of experimental results and simulated data, we demonstrate that as the memory window decreases: 1) The ferroelectric characteristic of Hf$_{0.5}$Zr$_{0.5}$O$_2$ is not degraded. 2) The trap density in the upper bandgap of the gate stacks increases. 3) The reason for memory window decrease is increased trapped electrons after program operation but not related to hole trapping/de-trapping. Our work is helpful to study the charge trapping behavior of FeFET and the related endurance fatigue process.

*Index Terms*—FeFET, Si, ferroelectric, doped HfO$_2$, endurance fatigue, charge trapping.

## I. INTRODUCTION

THE discovery of ferroelectric property in doped HfO$_2$ rejuvenates the FeFET research activity due to its complete compatibility with Si CMOS technology, scaling potential, and excellent retention property [1-4]. After ~10 years of intensive research, this CMOS-compatible material innovation enabled the demonstration of a FeFET technology scaled to the 28 nm and 22 nm nodes utilizing a conventional high-κ/metal gate technology [2, 5, 6]. Fast switching speed (≤100 ns), switching voltages in the range of 2.5~6 V, and 10-year data retention have been demonstrated for ferroelectric HfO$_2$ [7]. However, the endurance characteristic of Si FeFET has been limited to be $10^4$~$10^5$ cycles during the period of 2011~Feb. 2021 [8-19]. Very recently, significantly improved endurance of >$10^{10}$ cycles is reported by using high-κ Si$_3$N$_4$ interlayr in replacement of SiO$_2$ [20]. Meanwhile, the Hf$_{0.5}$Zr$_{0.5}$O$_2$ based Si FinFET with SiO$_x$ interlayer also shows high endurance of >$10^{10}$ cycles [21]. Generally, the endurance of >$10^{14}$ cycles is preferred for the industry's mass production [7]. Thus the Si FeFET endurance is still a key urgent challenge.

The physical origin of endurance fatigue is considered as charge trapping and trap generation [9, 10, 22]. Especially the charge trapping and de-trapping behaviors play dramatic roles [23]. During the program/erase (PGM/ERS) process of doped HfO$_2$ based Si FeFET, the remnant polarization is generally 10~30 μC/cm$^2$. This results in that the electric field across the interfacial layer (SiO$_2$) (30~90 MV/cm) exceeds its breakdown field (<15 MV/cm), which causes significant charge trapping/de-trapping. Thus, the techniques of quantitatively extracting/characterizing the charge trapping amount are rather essential to understand the charge trapping behavior, as well as its role on endurance fatigue.

Generally, there are two kinds of methods to quantitatively evaluate the trapped charges. One kind is directly measuring the trapped charges [24, 25]. This method evaluates the trapped charges by subtracting the substrate charge change from the total charges influent into the substrate. The trapped charges are found to be ~$10^{14}$ cm$^{-2}$ [24, 25]. The other kind is by tracing the threshold voltage ($V_{th}$) shift during the stress and release processes [17, 26-28], similar to the bias temperature instability (BTI) measurement. This method evaluates the trapped charges by considering that the trapped charges induce the $V_{th}$ shift. The trap concentration is about $10^{20}$ cm$^{-3}$, and the trap energy band is localized at ~1.8 to 2 eV below the conduction band minimum (CBM) of Hf$_{0.5}$Zr$_{0.5}$O$_2$. In addition, there are several other methods to probe the traps, such as low-frequency noise [15, 29], atomic force microscope (AFM) [30], Kelvin probe force microscope (KPFM) [30], trap spectroscopy by charge injection and sensing (TSCIS) [27], and exhaustive photo depopulation spectrum (EPDS) [27]. Even though the above methods have been proposed to investigate the charge trapping

Manuscript received June 30, 2021. This work was supported by the National Natural Science Foundation of China under Grant No. 61904199 and 61904193, and in part by the Open Research Project Fund of State Key Laboratory of ASIC and System under Grant No. KVH1233021. (Corresponding authors: Hao Xu, Jinjuan Xiang, Junshuai Chai)

Shujing Zhao, Fengbin Tian, Hao Xu, Jinjuan Xiang, Tingting Li, Junshuai Chai, Jiahui Duan, Kai Han, Xiaolei Wang, Wenwu Wang, and Tianchun Ye are with Key Laboratory of Microelectronics Devices and Integrated Technology, Institute of microelectronics, Chinese academy of sciences, Beijing 100029, China. The authors are also with University of Chinese Academy of Sciences, Beijing 100049, China (xuhao@ime.ac.cn; xiangjinjuan@ime.ac.cn; chaijunshuai@ime.ac.cn).

Kai Han is with Weifang University, Shandong 261061, China.

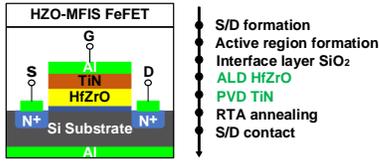

Fig. 1. Schematic of Si FeFET structure and its fabrication process.

behavior, the detailed mechanism of charging trapping induced endurance fatigue is still unclear. Generally, the endurance failure is ascribed to increased trapped charges with increasing PGM/ERS cycles [9, 10, 22, 25, 28]. On the contract, Ichihara *et al.* from Kioxia Corporation directly measured the trapped charges and found that trapped charges decreased with increasing PGM/ERS cycles [24]. Therefore, it is controversial whether the endurance failure is due to the increase of charge trapping. And the detailed physical picture of endurance fatigue by charge trapping is still unclear and consequently needs clarification.

In this work, we investigate the charge trapping during endurance fatigue of FeFET with TiN/Hf$_{0.5}$Zr$_{0.5}$O$_2$/SiO$_2$/Si (MFIS) gate structure. We propose a method of experimentally quasi-directly extracting the number of trapped charges during the memory operation. The amount of trapped charges increases during the endurance fatigue process. This is the first time that the trapped charges are directly experimentally extracted and verified to increase during endurance fatigue. Moreover, we model the interplay between the trapped charges and ferroelectric polarization switching during the endurance fatigue and give a detailed physical picture of charge trapping induced endurance fatigue.

## II. EXPERIMENTAL AND MODEL

### A. FeFET fabrication and electrical measurements

Fig. 1 shows a schematic diagram of the Si FeFET structure and its fabrication process. The FeFET was fabricated by the gate-last process. Firstly, the source and drain regions were formed by the As ion implantation for nMOSFET using the energy of 40 keV with a dose of 4×10$^{15}$ cm$^{-2}$, followed by annealing at 1050 °C for 5 s in N$_2$ atmosphere. Then the gate stack was grown. After diluted-HF clean, a 0.7 nm SiO$_2$ interfacial layer was grown by ozone oxidation at 300 °C in the atomic layer deposition (ALD) chamber. Then 9 nm Hf$_{0.5}$Zr$_{0.5}$O$_2$ was grown by ALD at 300 °C. The Hf precursor, Zr precursor, and O source were tetrakis-(ethylmethylamino)-hafnium (TEMA-Hf), tetrakis-(ethylmethylamino)-zirconium (TEMA-Zr), and H$_2$O, respectively. The 10 nm TiN and 75 nm W were grown by sputtering. Then ferroelectric phase crystallization was achieved by 550 °C for 60 s in N$_2$ to form orthorhombic phase. After the gate formation, the source/drain contacts were defined by lithography, and TiN/Al was used as contact metal. Finally, forming gas annealing at 450 °C in 5%-H$_2$/95%-N$_2$ was performed. In addition, the capacitor of TiN/9 nm Hf$_{0.5}$Zr$_{0.5}$O$_2$/TiN was also fabricated with the same growth condition as the FeFET.

The transfer characteristics $I_d$-$V_g$ and gate capacitance ($C_g$) were measured by Keysight B1500A. The polarization hysteresis loop was measured by the Radiant Precision LC ferroelectric tester. For the MFM capacitor, a triangle wave with ±3 V voltage amplitude at 1 kHz was used for hysteresis loop measurement. A square wave of ±3V voltage amplitude at 1 ms width was used for the endurance measurement. For the FeFET, AC small-signal at 100 kHz was used to measure $C_g$-$V_g$. A square wave of ±4.5 V voltage amplitude at 200 μs width was used for the endurance measurement. The threshold voltage was extracted by the linear extrapolation method.

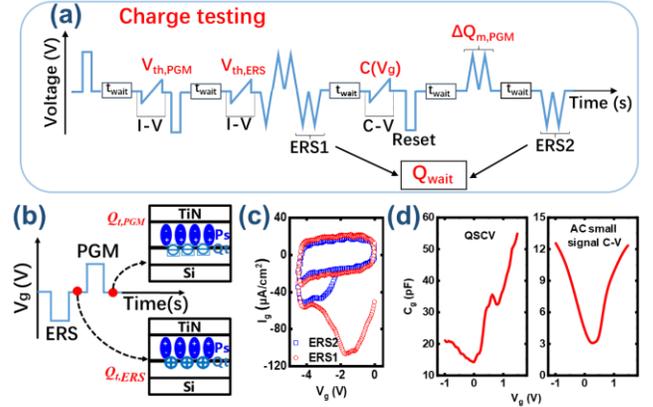

Fig. 2. (a) The method of measuring trapped charges. (b) The typical PGM/ERS operation of FeFET. (c) The measured gate current during the ERS1 and ERS2 processes. (d) Comparison between QSCV and high-frequency small AC signal C-V.

### B. Method of measuring trapped charges

Fig. 2(a)-(b) show the method of measuring trapped charges and the typical PGM/ERS operation of FeFET. Generally, electrons are trapped into the gate stack after PGM pulse, and here the trapped charge amount is denoted as $Q_{t,PGM}$. Holes are trapped after ERS pulse, and the trapped charge amount is denoted as $Q_{t,ERS}$. In this work, we experimentally measure the difference between the $Q_{t,PGM}$ and $Q_{t,ERS}$, i.e., $\Delta Q_t = Q_{t,PGM} - Q_{t,ERS}$, which represents the charge trapping behavior.

Next, we give the details of measuring $\Delta Q_t$. For the FeFET device, there are three kinds of free charges, including charges in the metal gate ($Q_m$), Si substrate ($Q_{Si}$) and trapped charges ($Q_t$). Similar to the definition of $\Delta Q_t$, we can define the $\Delta Q_m$ and $\Delta Q_{Si}$ correspondingly. According to the charge neutrality condition, we can obtain

$$\Delta Q_t + \Delta Q_m + \Delta Q_{Si} = 0 \quad (1)$$

Thus the $\Delta Q_t$ can be obtained after knowing the $\Delta Q_m$ and $\Delta Q_{Si}$.

The $\Delta Q_m$ is experimentally measured as shown in Fig. 2(a). After the negative pulse, two consecutive positive triangular waves are applied. Then the charge change on the metal gate is obtained by integrating the gate current, denoted as $\Delta Q_{m,PGM}$. It should be noted that the second triangular wave is used to eliminate the gate leakage, which is similar to the positive-up-negative-down (PUND) method. During our experimental measurement, we found that an unstable $Q_m$ component appears after the positive pulse, that is, part of $Q_m$ is released. Thus it is necessary to evaluate the amount of unstable $Q_m$ (denoted as $Q_{wait}$) to exactly obtain the $Q_t$. The *direct* measurement of $Q_{wait}$ needs a current measurement with a sampling time of fewer than 1 μs, together with a whole measurement time of more than 10 s simultaneously. Thus the *direct* measurement of $Q_{wait}$ is rather difficult. Here we measure the $Q_{wait}$ using the following alternative method as shown in Fig.

2(a). Firstly, after the positive triangular waves, two consecutive negative triangular waves (denoted as ERS1) are applied without waiting time. The charge change on the metal gate during this process ($\Delta Q_{m,ERS1}$) can be obtained by integrating the external current during the ERS1 process. Secondly, after the positive triangular wave and 100 s waiting time, two consecutive negative triangular waves (denoted as ERS2) are applied. The charge change on the metal gate during this process ($\Delta Q_{m,ERS2}$) can be obtained by integrating the external current during the ERS2 process. Here the waiting time of 100 s is used to ensure that the unstable charges are fully released. Fig. 2(c) shows the measured current during the above two processes. It can be seen that a current peak disappears for the ERS2 process compared with the ERS1 process. This indicates that the $Q_{wait}$ are released during the 100 s waiting time. Finally, the $Q_{wait}$ are determined as

$$Q_{wait} = \Delta Q_{m,ERS2} - \Delta Q_{m,ERS1} \tag{2}$$

Then the stable component ($\Delta Q_m$) can be obtained as

$$\Delta Q_m = \Delta Q_{m,PGM} - Q_{wait} \tag{3}$$

Then the $\Delta Q_{Si}$ is experimentally measured. After the PGM pulse, the substrate charge magnitude $Q_{Si,PGM}$ can be obtained if the $V_{th}$ and $C_g$-$V_g$ are known as

$$Q_{Si,PGM} = \int_0^{V_{th,PGM}} C_{g,PGM}(V_g) dV_g \tag{4}$$

where $C_{g,PGM}$ means the gate capacitance after PGM pulse. Similarly, after the ERS pulse, the substrate charge magnitude $Q_{Si,ERS}$ can be obtained as

$$Q_{Si,ERS} = \int_0^{V_{th,ERS}} C_{g,ERS}(V_g) dV_g \tag{5}$$

Then the $\Delta Q_{Si}$ can be obtained as

$$\begin{aligned}\Delta Q_{Si} &= Q_{Si,PGM} - Q_{Si,ERS}\\ &= \int_0^{V_{th,PGM}} C_{g,PGM}(V_g)dV_g - \int_0^{V_{th,ERS}} C_{g,ERS}(V_g)dV_g \\ &= \int_{\Delta V_{th}}^{V_{th,ERS}} C_{g,ERS}(V_g)dV_g - \int_0^{V_{th,ERS}} C_{g,ERS}(V_g)dV_g \\ &= -\int_0^{\Delta V_{th}} C_{g,ERS}(V_g)dV_g\end{aligned} \tag{6}$$

Finally, the $\Delta Q_t$ can be obtained from (1).

In addition, the spontaneous polarization change (denoted as $\Delta P_S$) corresponding to $\Delta Q_t$ process can be evaluated. The $V_{th}$ shift due to $\Delta Q_t$ and $\Delta P_S$ can be given as

$$\Delta V_{th} = -\frac{\Delta Q_t + \Delta P_S}{C_{FE}} = -\frac{\Delta Q_t + \Delta P_S}{\varepsilon_0 \varepsilon_{FE}} d_{FE} \tag{7}$$

where $C_{FE}$ means the ferroelectric capacitance. The $\varepsilon_0$ and $\varepsilon_{FE}$ represent vacuum and relative dielectric constants of ferroelectric, respectively. $d_{FE}$ means the physical thickness of ferroelectric. Then the $\Delta P_S$ can be known as

$$\Delta P_S = \Delta Q_t - \frac{\varepsilon_0 \varepsilon_{FE}}{d_{FE}} \Delta V_{th} \tag{8}$$

It should be noted that our method is based on the work by Ichihara *et al.* from Kioxia Corporation [24]. Their method can use one-time current measurement to extract the $\Delta Q_t$. However, there are two issues about their method. Firstly, their method overestimates the $\Delta Q_{Si}$. They measured the $C_g$-$V_g$ by the direct current method, in other words, quasi-static $C_g$-$V_g$ (QSCV). Thus the measured $C_g$ contains the contribution from ferroelectric spontaneous polarization switching. The $C_g$ and resulted $\Delta Q_{Si}$ are overestimated. While in our work, we use the high-frequency small AC signal to measure the $C_g$-$V_g$. This can significantly suppress the ferroelectric contribution. Moreover, we also experimentally compared the two methods. We found that the $\Delta Q_{Si}$ obtained by the QSCV method is 71.1% larger than that by the AC method, the $C_g$-$V_g$ curve shown in Fig. 2(d). Similar results have been reported [17], which further supports the above discussion. Secondly, their method results in overestimated $\Delta Q_m$. Their method measured the $Q_{wait}$ with an additional triangle wave for $I_d$-$V_g$ test. This induced additional electron trapping and consequently underestimated $Q_{wait}$. We found that the $\Delta Q_m$ obtained with an additional triangle wave is 5.6% larger than that without an additional triangle wave (not shown here). While in our method, we independently measure $Q_{wait}$ and $V_{th,PGM}$ to remove the additional triangle wave, and can more accurately evaluate the $\Delta Q_m$.

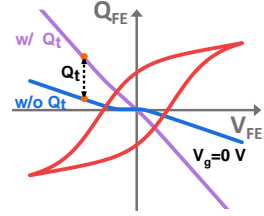

Fig. 3. The effect of charge trapping on loadline.

### C. Modeling of MFIS gate stack with charge trapping

We model voltage and charge distribution across the MFIS gate stack by using the polarization-electric field curve ($P$-$E_{FE}$) of ferroelectric and loadline curve, based on our previous work [31]. The hysteresis loop of the ferroelectric is described by the Miller model. The loadline is given as

$$\begin{aligned}V_{FE} &= V_g - \psi_S - V_{IL} = V_g - \psi_S + \frac{Q_{Si}(\psi_S)}{\varepsilon_0 \varepsilon_{IL}} t_{IL}\\ &= V_g - \psi_S - \frac{Q_m(\psi_S) + Q_t(\psi_S)}{\varepsilon_0 \varepsilon_{IL}} t_{IL}\end{aligned} \tag{9}$$

where the $\varepsilon_0$ and $\varepsilon_{IL}$ represent vacuum and relative dielectric constants of the interlayer, respectively. $t_{IL}$ means the physical thickness of the interlayer. The term $Q(\psi_S)$ means that the $Q$ is a function of $\psi_S$. The (9) has used the (1). From the (9), we can obtain a $V_{FE}$ value for each $\psi_S$ at a given $V_g$. Then the $Q_m(\psi_S)$ vs. $V_{FE}$ curve, i.e., the loadline, can be obtained.

We discuss the relationship between the $Q_t$ and $\psi_S$. The charge trapping/de-trapping occurs during the PGM/ERS process. The occupation of traps is influenced by the relative position between the trap energy levels and Fermi level. Thus, as $\psi_S$ and $V_{IL}$ increases, the negative trapped charges $Q_t$ becomes larger. Similarly, as $\psi_S$ and $V_{IL}$ decreases, the positive trapped charges $Q_t$ becomes larger.

We discuss the role of $Q_t(\psi_S)$ on the loadline. The introduction of $Q_t$ will cause the loadline to become steeper. The reason is given as follows. Here we explain the case of negative charge trapping. At a given $V_g$, for the same $V_{FE}$ condition, the $\psi_S$ is identical no matter of with and without negative trapped charges. With identical $\psi_S$, $Q_m$ increases due to the negative trapped charges base on (1). Thus the point on the loadline shifts upwards when negative charges are trapped, as shown in Fig. 3. With decreasing the $V_{FE}$, more negative charges are trapped, resulting in a larger upward shift. Therefore the loadline becomes steeper. For the case of positive charge trapping the conclusion is same.

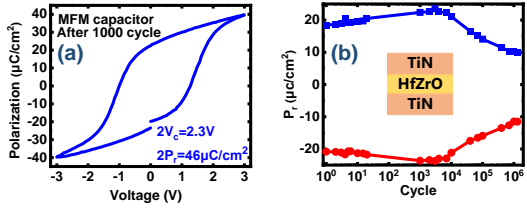

Fig. 4. (a) P-V of the Hf$_{0.5}$Zr$_{0.5}$O$_2$ MFM capacitor after 10$^3$ PGM/ERS cycles and (b) the endurance process.

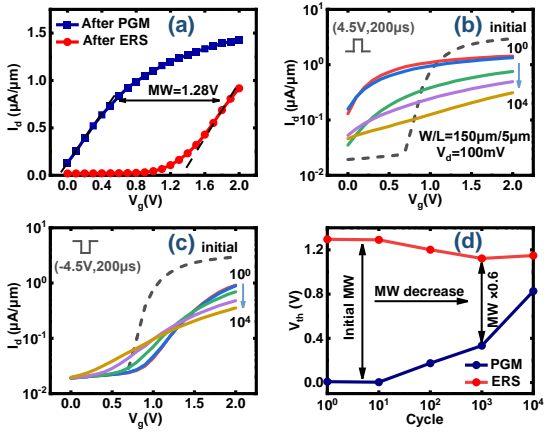

Fig. 5. (a) $I_d$-$V_g$ curve after (a) wake up, and after (b) PGM and (c) ERS during the endurance process. (d) The MW. This device is woken up after 1s, ±4.5V pulse.

Based on the model we can calculate voltage and charge distribution across the MFIS gate stack, especially the $\Delta Q_t$ and $\Delta Q_m$, and compare them with the experimental results.

### III. RESULTS AND DISCUSSION

#### A. Polarization characteristics of Hf$_{0.5}$Zr$_{0.5}$O$_2$ ferroelectric

Fig. 4(a) shows the P-V curve of the Hf$_{0.5}$Zr$_{0.5}$O$_2$ MFM capacitor after 10$^3$ PGM/ERS cycles. The remnant polarization 2P$_r$ is about 46 μC/cm$^2$, and coercive voltage 2V$_c$ is about 2.3 V (coercive field 2E$_c$ ~ 2.6 MV/cm). Fig. 4(b) shows the endurance process. When the cycle number increases from 10$^0$ to 10$^4$, the 2P$_r$ increases from 36 to 46 μC/cm$^2$. From 10$^4$ to 10$^6$, 2P$_r$ decreases. Thus the ferroelectric characteristic of the MFM capacitor does not degrade until 10$^4$ cycles.

#### B. Endurance fatigue of Hf$_{0.5}$Zr$_{0.5}$O$_2$ FeFET

Fig. 5(a) shows the $I_d$-$V_g$ curve after wake up. The memory window (MW) is 1.28 V. Fig. 5(b) and (c) show $I_d$-$V_g$ curves during the endurance process. The $V_{th}$ can be extracted and the MW is shown in Fig. 5(d). The MW decreases with increasing the PGM/ERS cycle. After 10$^3$ cycles, the MW is reduced to be 60%. After 10$^4$ cycles, the MW is reduced to be 20%. After 10$^5$ cycles, the MW disappeared. From the endurance difference between the MFM capacitor and FeFET, we can conclude that the charge trapping plays a dramatic role on the FeFET, which is consistent with the reported results [9, 10, 16, 22, 25, 26].

#### C. Quantitative characterization of charge trapping during the endurance fatigue process

Firstly, we experimentally investigate the $\Delta Q_m$. Fig. 6(a)

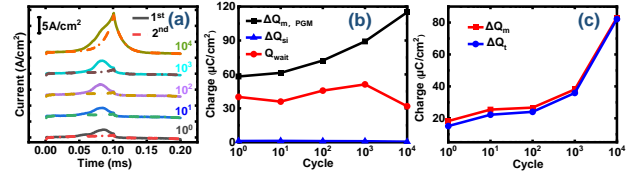

Fig. 6. (a) External current during two consecutive positive triangular waves. (b) $\Delta Q_{m,PGM}$, $Q_{wait}$ and $\Delta Q_{Si}$ during endurance fatigue. (c) $\Delta Q_m$ and $\Delta Q_t$ during endurance fatigue.

shows external current during two consecutive positive triangular waves (see Fig. 2(a)). The current corresponding to the 1$^{st}$ triangular wave is larger than the 2$^{nd}$ one. This indicates the ferroelectric spontaneous polarization reversal. Then the $\Delta Q_{m,PGM}$ is obtained as shown in Fig. 6(b). The $\Delta Q_{m,PGM}$ increases from 58 to 114 μC/cm$^2$ as the PGM/ERS cycle increases. The $Q_{wait}$ is also shown in Fig. 6(b). The $Q_{wait}$ does not change significantly and is about 40 μC/cm$^2$. It indicates that the trap density corresponding to unstable charge trapping is not changed significantly. Finally, the $\Delta Q_m$ is obtained as shown in Fig. 6(c). The $\Delta Q_m$ increases from 18 to 83 μC/cm$^2$ as the PGM/ERS cycle increases. The leakage current across the gate stack increases during the endurance fatigue process as shown in Fig. 6(a). The maximum leakage increases from 5 μA after 10$^0$ cycle to 87μA after 10$^4$ cycles. This indicates that significant defect generation process occurs.

Secondly, we experimentally investigate the $\Delta Q_{Si}$. The $\Delta Q_{Si}$ can be obtained based on (6), and shown in Fig. 6(b). The $\Delta Q_{Si}$ decreases from 3.15 to 0.93 μC/cm$^2$ due to MW reduction.

Finally, we experimentally investigate the $\Delta Q_t$, as shown in Fig. 6(c). The $\Delta Q_t$ significantly increases, from 17 μC/cm$^2$ after 10$^0$ cycle to 80 μC/cm$^2$ after 10$^4$ cycles. Considering that the surface potential difference between the two states after PGM and ERS pulse decreases during the endurance fatigue, the increase in $\Delta Q_t$ is caused by the increase of density of the stable trap of the gate stack. Furthermore, we conclude that the endurance fatigue is originated from the increase of $\Delta Q_t$.

It should be noted that our work directly verifies that the trapped charges increase during endurance fatigue for the first time. For the method of evaluating the trapped charges by tracing the $V_{th}$ [17, 26-28], it cannot directly measure the trapped charges. Considering that the endurance fatigue could also originate from the fatigue of ferroelectric, this method can only speculatively ascribe endurance fatigue to the charge trapping behavior. For the method of directly measuring the trapped charges, it found that the trapped charges decreased during endurance fatigue [24]. Thus charge trapping cannot be considered as the origin based on their results. Our method evaluates the charge trapping behavior experimentally and directly. We find that the trapped charges increase during endurance fatigue. Thus it is the first time to credibly report the enhancement of charge trapping during endurance fatigue.

#### D. Simulation results

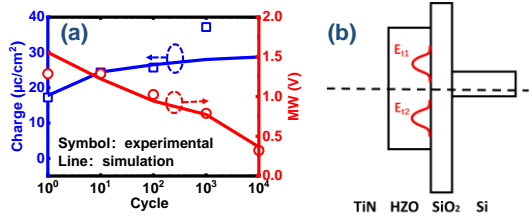

Fig. 7. (a) Experimental and simulation results of the MW and $\Delta Q_t$. (b) Schematic of the traps at FE/DE interface.

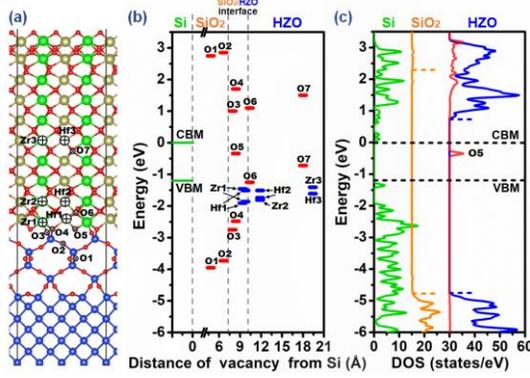

Fig.8. (a) The $Hf_{0.5}Zr_{0.5}O_2/SiO_2/$ Si gate stack model. The "⊕" and "⊗" represent metal and O atom vacancies, respectively. Several physical positions of metal or O vacancy are considered. (b) The energy level distribution of vacancy defects. (c) The projected density of states for the case of O5 vacancy defect.

To further understand how the charge trapping degrades the endurance characteristic, we simulate the relationship between the trapped charge and the endurance fatigue process. Fig. 7(a) shows the experimental results about the MW and trapped charges $\Delta Q_t$, together with the simulated data. It can be seen that the experimental and simulated data are well consistent. This indicates the validity of the modeling. The parameters in our modeling are determined as follows. The ferroelectric characteristic parameters (remnant polarization (23 μC/cm$^2$), saturated spontaneous polarization (30.2 μC/cm$^2$), and coercive field (1.28 MV/cm)) were determined by fitting the *P-V* characteristics of the MFM structure. In addition, the ferroelectric characteristic parameters were considered unchanged during the endurance fatigue of FeFET. This is supported by Fig. 4(b), where no ferroelectric degradation appears until $10^4$ cycles. Then these parameters were introduced into the modeling to obtain information about the trap distribution and generation. Based on our consistent modeling of experimental results as shown in Fig. 7(a), we find that the traps at $Hf_{0.5}Zr_{0.5}O_2/SiO_2$ interface are localized with two energy levels, as shown in Fig. 7(b). One energy position is 0.36 eV above the CBM of the Si substrate (denoted as $E_{t1}$), and the other is 0.76 eV below the valence band minimum (VBM) of the Si substrate (denoted as $E_{t2}$). The trap position based on our work is consistent with published results [17, 27]. Moreover, the density of $E_{t1}$ increases from $1 \times 10^{14}$ cm$^{-2}$ after 1 PGM/ERS cycle to $3.6 \times 10^{14}$ cm$^{-2}$ after $10^4$ cycles. Whereas the density of $E_{t2}$ is nearly unchanged during the endurance fatigue ($\sim 1 \times 10^{14}$ cm$^{-2}$). This is consistent with Fig. 5(b), where the $V_{th}$ after ERS operation is nearly unchanged and the $V_{th}$ after PGM operation increases with increasing endurance cycles. Furthermore, to determine the physical origin of traps, we construct a $Hf_{0.5}Zr_{0.5}O_2/SiO_2/Si$ gate stack structure (see Fig. 8(a)) and investigate the distribution of trap energy levels by calculating the projected density of state (DOS) (see Fig. 8(b) and 8(c)) based on *ab initio* calculations employing the HSE06 method [32-34]. It can be seen that the trap energy level near the CBM is mainly contributed to the two-coordinated O atom vacancy at the $Hf_{0.5}Zr_{0.5}O_2/SiO_2$ interface, while the trap energy level at about 0.65 eV below the VBM is mainly from the metal atom vacancies near the $Hf_{0.5}Zr_{0.5}O_2/SiO_2$ interface (Hf1, Hf2, Zr1, Zr2) and in the $Hf_{0.5}Zr_{0.5}O_2$ bulk (Hf3). Thus the $E_{t1}$ is considered to originate from two-coordinated O atom vacancy at the $Hf_{0.5}Zr_{0.5}O_2/SiO_2$ interface, and the $E_{t2}$ is considered to originate from metal vacancles at the $Hf_{0.5}Zr_{0.5}O_2/SiO_2$ interface and in $Hf_{0.5}Zr_{0.5}O_2$ bulk.

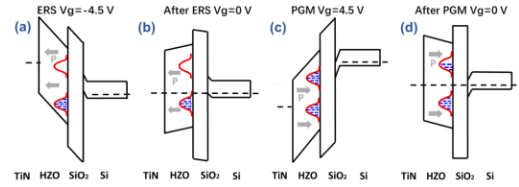

Fig. 9. Energy band diagram (a) at ERS, (b) after ERS, (c) at PGM, and (d) after PGM.

Then we discuss the detailed physical picture of charge trapping induced endurance fatigue, based on our modeling. Fig. 9(a) shows the energy band diagram *at* ERS condition. The Fermi level is localized above the $E_{t2}$ trap bands and below the $E_{t1}$ trap bands. Thus the $E_{t2}$ trap bands are occupied by electrons and the $E_{t1}$ trap bands are empty. In addition, the $Hf_{0.5}Zr_{0.5}O_2$ is negatively polarized. Fig. 9(b) shows the energy band diagram *after* ERS condition. The Fermi level is localized between the $E_{t1}$ and $E_{t2}$ trap bands. Thus the charge trapping is nearly identical between Fig. 9(a) and 9(b). In addition, the $Hf_{0.5}Zr_{0.5}O_2$ is negatively polarized. Moreover, our modeling finds that the energy band diagram for the ERS operation is similar during the fatigue process. Therefore, the $V_{th}$ is nearly uncharged for the ERS operation. For the PGM operation, Fig. 9(c) shows the energy band diagram *at* PGM condition. The Fermi level is localized above the $E_{t1}$ trap bands. Thus both the $E_{t1}$ and $E_{t2}$ trap bands are occupied by electrons. In addition, the $Hf_{0.5}Zr_{0.5}O_2$ is positively polarized. Fig. 9(d) shows the energy band diagram *after* PGM. The Fermi level is localized between the $E_{t1}$ and $E_{t2}$ trap bands. However, the trapped electrons in the $E_{t1}$ trap bands cannot fully be de-trapped. In addition, the $Hf_{0.5}Zr_{0.5}O_2$ is positively polarized. Moreover, our modeling finds that the $E_{t1}$ traps increases during endurance fatigue. Therefore, the $V_{th}$ increases during endurance fatigue. In conclusion, the endurance fatigue of $Hf_{0.5}Zr_{0.5}O_2$ Si FeFET is due to the increased charge trapping in the upper trap bands.

## IV. CONCLUSION

This work presents a method to quantitatively characterize charge trapping of FeFET with TiN/HfZrO/SiO$_2$/Si gate structure, and studies the detailed physical mechanism of how the charge trapping causes fatigue degradation. We experimentally verify that the amount of trapped charges increases during the endurance fatigue process. This is the first time that the trapped charges are directly experimentally extracted and verified to increases during endurance fatigue.

Moreover, we model the charge trapping and its role in endurance fatigue. We find that the endurance fatigue of FeFET is not due to the fatigue of $Hf_{0.5}Zr_{0.5}O_2$ ferroelectric property, but the increased trap density in the upper energy bandgap. Our work gives a detailed physical picture of endurance fatigue by charge trapping. Our work clearly verifies the charge trapping as the physical origin of endurance fatigue.